\documentstyle[twoside,fleqn,espcrc2]{article}

\newcommand{\bfk}{\mbox{\boldmath $k$}} 
\newcommand{\bfs}{\mbox{\boldmath $s$}} 
\newcommand{\bfp}{\mbox{\boldmath $p$}} 
\newcommand{\bfP}{\mbox{\boldmath $P$}} 
\newcommand{\pup}{p^\uparrow}
\newcommand{\qup}{q^\uparrow}
\newcommand{\hup}{h^\uparrow}
\newcommand{\bup}{b^\uparrow}
\newcommand{\pdown}{p^\downarrow}
\newcommand{\qdown}{q^\downarrow}
\newcommand{\hdown}{h^\downarrow}
\newcommand{\be}{\begin{equation}}
\newcommand{\ee}{\end{equation}}
\newcommand{\bea}{\begin{eqnarray}}
\newcommand{\eea}{\end{eqnarray}}
 
\newcommand{\gsim}{\raisebox{-4pt}{$\, \stackrel{\textstyle >}{\sim} \,$}}

\newcommand{\AmS}{{\protect\the\textfont2
  A\kern-.1667em\lower.5ex\hbox{M}\kern-.125emS}}

\title{Origins of single transverse spin asymmetries}

\author{M. Anselmino
       \address{Dipartimento di Fisica Teorica, Universit\`a di Torino and  
      INFN, Sezione di Torino, Via P. Giuria 1, I-10125 Torino, Italy},
        M. Boglione
       \address{Department of Physics, University of Durham, Science 
       Laboratories, South Road, Durham DH11 3LE, United Kingdom}, 
       U. D'Alesio$^{\,\>\rm c}$ and F. Murgia
       \address{Istituto Nazionale di Fisica Nucleare, Sezione di Cagliari 
       and Dipartimento di Fisica, Universit\`a di Cagliari,
       C.P. 170, I-09042 Monserrato (CA), Italy}}

\begin{document}

\begin{abstract}
Single transverse spin asymmetries in $p\,p \to \pi\,X$ and 
$\ell\,p \to \ell\,\pi\,X$ processes have been observed; their possible 
origins and connections are investigated. A phenomenological description 
within a pQCD generalized factorization scheme is discussed.  
\vspace{-1pc}
\end{abstract}

\maketitle

\section{Introduction and data}

Several spin asymmetries in processes involving only one transversely 
polarized hadron have been observed experimentally; these, when occurring
in a kinematical region where pQCD should be applicable, pose a severe 
challenge to a correct phenomenological description, as single spin 
asymmetries are negligible in the elementary interactions, due to chirality
conservation of QCD and QED dynamics. Single transverse spin asymmetries
are related to helicity flip amplitudes and to relative phases, both of which
are absent in the perturbative, leading order interactions of quarks, 
gluons and photons. 

Single transverse spin asymmetries are then sensitive to higher twist 
contributions, or non perturbative effects in the long distance physics,
and are expected to vanish in the truly asymptotic, high energy, large 
$Q^2$ (or $p_T$) regions, where leading twist and parton collinear 
configurations dominate in the QCD factorization scheme. Most data are not 
from that region yet, and the investigation of single spin effects, both in 
experiments and theories, is bound to be rich of unexpected and new results. 

We discuss here two kinds of single spin asymmetries, occurring in  
$p \, p \to \pi \, X$ and $\ell \, p \to \ell \, \pi \, X$ processes
with a polarized initial proton; the cross-sections depend on the 
proton spin orientation, whether $\uparrow$ or $\downarrow$ with respect 
to the production plane, giving origin to an asymmetry,  
\be
A_N \equiv \frac{d\sigma^\uparrow - d\sigma^\downarrow}
{d\sigma^\uparrow + d\sigma^\downarrow} \> \cdot
\label{spa}
\ee

Large values of $A_N$ for the first process have been measured already 
several years ago \cite{e704}; more recently also $A_N$ for the  
semi-inclusive DIS process has been found to be different from zero
\cite{her,smc}. 
%They are shown in Figs. 1 and 2.      
 
The two asymmetries are interesting in many respects: they might have a 
common origin -- the so called Collins quark fragmentation function 
\cite{col} -- and are going to be measured again very soon, respectively at 
RHIC and by HERMES collaboration, with an upgraded transversely 
polarized proton target. It is then appropriate to have a discussion 
about the possible origins of $A_N$ and to develop a phenomenological
approach towards their description and prediction. 
 
Among the attempted explanations of $A_N$ observed in E704 experiment
there are generalizations of the QCD factorization theorem with the
inclusion of higher twist correlation functions \cite{ekt,qs,kk}, or with the 
inclusion of intrinsic $\bfk_\perp$ and spin dependences in distribution
\cite{siv,noi1,bm,dan} and fragmentation \cite{col,bm,noi2,bl,suz} functions; 
there are also some semi-classical approaches based on introduction of 
quark orbital angular momentum \cite{bztm,bzt}. A review paper on the 
subject can be found in Ref. \cite{bzt}.
We consider here only the approaches which are based on QCD dynamics, through
a generalization of the factorization scheme, according to which, at leading 
twist and with collinear configurations, the cross-section for the  
process $p\,p \to \pi\,X$ can be written as the usual convolution,
\be
d\sigma = \sum_{a,b,c} f_{a/p} \otimes f_{b/p} \otimes
d\hat\sigma^{ab \to c \dots} \otimes D_{\pi/c} \>,
\label{lt}
\ee
in terms of distribution and fragmentation functions and pQCD partonic 
interactions. This simple approach, however, predicts negligible single spin
asymmetries and higher order contributions have to be taken into account.      
  
\section{Higher twist parton correlations}

In the approach of Ref. \cite{qs} Eq. (\ref{lt}) is generalized -- and proven
to hold -- with the introduction of higher twist contributions to distribution 
or fragmentation functions. Schematically it reads: 
\bea  
d\sigma^\uparrow - d\sigma^\downarrow \!\!\!\! &=& \!\!\!\! \sum_{a,b,c} 
\Bigl\{ \Phi_{a/p}^{(3)} \otimes f_{b/p} \otimes \hat H \otimes D_{\pi/c} 
\nonumber \\
&+& \!\! h_1^{a/p} \otimes \Phi_{b/p}^{(3)} \otimes \hat H' 
\otimes D_{\pi/c} \\
&+& \!\! h_1^{a/p} \otimes f_{b/p} \otimes \hat H'' \otimes 
D_{\pi/c}^{(3)} \Bigr\}\>,
\nonumber 
\eea 
where the $\Phi^{(3)}$'s, $D^{(3)}$'s, are the higher twist correlations and
the $\hat H$'s denote the elementary interactions.
$h_1$ is the transversity distribution. 

The higher twist contributions are unknown, but some simple models can 
be introduced, for example
\bea
&& \!\!\!\!
\Phi^{(3)}_{a/p} \sim \int\frac{dy^-}{4\pi}\>  e^{ixp^+y^-} \langle p, \bfs_T|
\, \overline\psi_a(0) \gamma^+ \times \nonumber \\
&& \!\!\!\!\!\!\!\!\!\! \left[ \int \!\! dy_2^- \, 
\epsilon_{\rho\sigma\alpha\beta} 
\> s_T^\rho \, p^\alpha \, p'^{\beta} \, F^{\sigma+}(y_2)\right]
\psi_a(y^-) \, |p, \bfs_T \rangle \nonumber \\
&=& \!\! k_a \> C \> f_{a/p} \>. \label{phi3}
\eea

The above contribution depends on the initial nucleon momenta $p$ and $p'$,
on the transverse proton spin $\bfs_T$ and on some external gluonic field
$F^{\mu\nu}$; it involves transverse degrees of freedom of the partons
and it differs from the usual definition of the distribution functions
$f_{a/p}$ only by the insertion of the term in squared brackets.
This is the reason 
for the last line of Eq. (\ref{phi3}), where $C$ is a dimensional parameter 
and $k_a$ is respectively $+1$ and $-1$ for $u$ and $d$ quarks.

Such a simple model can reproduce the main features of the data \cite{e704} 
and some predictions for RHIC energy can be attempted \cite{qs}. 

\section{Intrinsic $\bfk_\perp$ in QCD factorization}

A somewhat analogous approach has been discussed in 
Refs.~\cite{siv,noi1,dan,noi2}; again, one starts from the leading twist,
collinear configuration scheme of Eq. (\ref{lt}), and generalizes it
with the inclusion of intrinsic transverse motion of partons in distribution
functions and hadrons in fragmentation processes.

The introduction of $\bfk_\perp$ and spin dependences opens up the way to
many possible spin effects; these can be summarized by the new functions:
\bea
\Delta^Nf_{q/\pup}  \!\!\! &\equiv& \!\!\!
\hat f_{q/\pup}(x, \bfk_{\perp})-\hat f_{q/\pdown}(x, \bfk_{\perp}) 
\label{delf1}\\
\!\!\! & = & \!\!\! 
\hat f_{q/\pup}(x, \bfk_{\perp})-\hat f_{q/\pup}(x, - \bfk_{\perp}) 
\nonumber \\
\Delta^Nf_{\qup/p}  \!\!\! &\equiv& \!\!\!
\hat f_{\qup/p}(x, \bfk_{\perp})-\hat f_{\qdown/p}(x, \bfk_{\perp}) 
\label{delf2}\\
\!\!\! & = & \!\!\! 
\hat f_{\qup/p}(x, \bfk_{\perp})-\hat f_{\qup/p}(x, - \bfk_{\perp})  
\nonumber \\
\Delta^N D_{h/\qup} \!\!\! &\equiv& \!\!\!
\hat D_{h/\qup}(z, \bfk_{\perp}) - \hat D_{h/\qdown}(z, \bfk_{\perp}) 
\label{deld1}\\
\!\!\! & = & \!\!\! 
\hat D_{h/\qup}(z, \bfk_{\perp})-\hat D_{h/\qup}(z, - \bfk_{\perp}) 
\nonumber \\
\Delta^N D_{\hup/q} \!\!\! &\equiv& \!\!\!
\hat D_{\hup/q}(z, \bfk_{\perp}) - \hat D_{\hdown/q}(z, \bfk_{\perp}) 
\label{deld2}\\
\!\!\! &=& \!\!\! 
\hat D_{\hup/q}(z, \bfk_{\perp})-\hat D_{\hup/q}(z, - \bfk_{\perp}) 
\>, \nonumber
\eea
which have a clear meaning if one pays attention to the arrows denoting the
polarized particles. Details can be found, for example, in Ref. \cite{vill}.
All the above functions vanish when $k_\perp=0$ and are na\"{\i}vely $T$-odd.
The ones in Eqs. (\ref{delf2}) and (\ref{deld1}) are chiral-odd, while the 
other two are chiral-even.   

Similar functions have been introduced in the literature with different 
notations: in particular there is a direct correspondence \cite{bm} between 
the above functions and the ones denoted, respectively, by: $f_{1T}^\perp$
\cite{dp}, $h_1^\perp$ \cite{dan}, $H_1^\perp$ and 
$D_{1T}^\perp$ \cite{dp,jac}.
The function in Eq. (\ref{deld1}) is the Collins function \cite{col},
while that in Eq. (\ref{delf1}) was first introduced by Sivers \cite{siv}.       
By inserting the new functions into Eq. (\ref{lt}), and keeping only 
leading terms in $k_\perp$, one obtains:
\bea
&& d\sigma^\uparrow - d\sigma^\downarrow = \sum_{a,b,c} 
\label {gen} \\ 
&\!\!\!\Bigl\{& \!\!\!\!\!
\Delta^Nf_{a/\pup} (\bfk_{\perp}) \otimes f_{b/p}
\otimes
d\hat\sigma(\bfk_{\perp}) \otimes D_{\pi/c} \nonumber \\
&\!\!\!+& \!\!\!\! h_1^{a/p} \otimes f_{b/p} \otimes
\Delta\hat\sigma(\bfk_\perp) \otimes \Delta^N D_{\pi/c}(\bfk_\perp) 
\nonumber\\  
&\!\!\!+& \!\!\!\! h_1^{a/p} \otimes \Delta^Nf_{\bup/p} (\bfk_{\perp}) \otimes
\Delta'\hat\sigma(\bfk_\perp) \otimes D_{\pi/c}(z) \Bigr\}\>, \nonumber
\eea
where the convolution now involves also a $\bfk_\perp$ integration
(we have explicitely shown the $\bfk_\perp$ dependences). The 
$\Delta\hat\sigma$'s denote polarized elementary interactions,
computable in pQCD. Notice that in the physical quantity above 
only products of even numbers of chiral-odd functions appear.

The above expression has been used to successfully fit the E704 data,
either with the Sivers effect only \cite{noi1} (second line of Eq. (\ref{gen}))
or the Collins effect only \cite{noi2,bl} (third line).
Some words of caution are necessary concerning the Sivers 
function $\Delta^Nf_{q/\pup}$, which is proportional to off-diagonal
(in helicity basis) expectation values of quark operators between
proton states \cite{col}: 
\be
\Delta^Nf_{a/\pup} \sim 
\langle p+|\, \overline\psi \gamma^+ \psi \, |p\, -\rangle \>.
\ee    

By exploiting the usual QCD parity and time-reversal properties for
free states one can prove the above quantity to be zero \cite{col}.
This might eliminate the Sivers function from the possible phenomenological
explanations of single spin asymmetries. However, Sivers effect might be
rescued by initial state interactions, or by a new and subtle interpretation
of time reversal properties, discussed in the talk by A. Drago \cite{dra}. 
  
\section{Fragmentation of polarized quarks}

The Collins function, Eq. (\ref{deld1}), can explain the E704 data on 
$p \, p \to \pi \, X$ single spin asymmetry \cite{noi2,bl}; are there other 
ways of accessing it, in order to get independent estimates of its size?

The answer to this question brings us to the azimuthal asymmetry observed 
by HERMES and SMC collaborations in semi-inclusive DIS \cite{her,smc}, 
$\ell \, p \to \ell \, \pi \, X$. Such asymmetries are directly related 
to the Collins function. In fact Eq. (\ref{deld1}) can be rewritten as:
\bea 
&& D_{h/\qup}(z, \bfk_\perp) = \hat D_{h/q}(z, k_\perp) \label{colfn} \\
&+& \!\!\!\! \frac 12 \> \Delta^ND_{h/\qup}(z, k_\perp) \> 
\frac{\bfP_q \cdot (\bfp_q \times \bfk_\perp)}
{|\bfp_q \times \bfk_\perp|}\>, \nonumber 
\eea
for a quark with momentum $\bfp_q$ and a transverse polarization 
vector $\bfP_q$ ($\bfp_q \cdot \bfP_q = 0$) which fragments into a hadron 
with momentum $\bfp_h = z\bfp_q + \bfk_\perp$ ($\bfp_q \cdot \bfk_\perp = 0$). 
$\hat D_{h/q}(z, k_\perp)$ is the unpolarized, $k_\perp$ dependent, 
fragmentation function. Parity invariance demands that the only component 
of the polarization vector which contributes to the spin dependent part
of $D$ is that perpendicular to the $q-h$ plane; in general one has:
\be
\bfP_q \cdot 
\frac{\bfp_q \times \bfk_\perp} {|\bfp_q \times \bfk_\perp|}
= P_q \sin\Phi_C \>, \label{colan}
\ee
where $P_q = |\bfP_q|$ and $\Phi_C$ is the Collins angle. When $P_q =1$ and 
$\bfP_q$ is perpendicular to the $q-h$ plane ($\bfP_q = \, \uparrow$, 
$-\bfP_q = \, \downarrow$) one has $P_q \sin\Phi_C = 1$. 

Eq. (\ref{colfn}) then implies the existence of a {\it quark analysing power}:
\be
A_q^h \equiv \frac {\hat D_{h/\qup} - \hat D_{h/\qdown}} 
{\hat D_{h/\qup} + \hat D_{h/\qdown}} = 
\frac{\Delta^ND_{h/\qup}}{2 \hat D_{h/q}} \>\cdot
\ee

The question about the possible size of $A_q^\pi$, as derived from the 
experimental data, has been addressed in Ref. \cite{noi3}. 

The asymmetry $A_N$, Eq. (\ref{spa}), for the 
$\ell \, p \to \ell \, \pi^+ \, X$ process, is given, at leading twist, by:
\be
A_N^{\pi^+} = \frac{4h_1^{u/p}}{4f_{u/p} +f_{\bar d/p}} \,
A_q^\pi \, \frac{2(1-y)}{1 + (1-y)^2} \, \sin\Phi_C \label{azi}
\ee
in terms of the usual DIS variables. Similar expressions hold for 
$\pi^-$ and $\pi^0$. Notice that $A_N^\pi$ depends on the product of the 
Collins function and the transversity distribution (as required by chirality
conservation): each of them depends on different variables ($z$ for $A_q^\pi$
and $x$ for $h_1$) and, in principle, accurate measurements could provide
access to both functions. 

Eq. (\ref{azi}) should be compared with data in kinematical regions such 
that leading twist contributions are dominant; this is not the case for 
HERMES experiment, as they have protons with a longitudinal (with respect 
to the lepton direction) polarization, whose transverse (with respect to
$\gamma^*$ direction) component is depressed by a $1/Q$ factor, making it
effectively a higher twist contribution. The situation is better with SMC 
data, obtained with transversely polarized protons, although their results
are still preliminary (and very likely will remain such); anyway, they 
have \cite{smc,noi3}:
\be
A_N^{\pi^+} \simeq -(0.10 \pm 0.06) \> \sin\Phi_C \>. \label{dat}
\ee

By comparing Eqs. (\ref{azi}) and (\ref{dat}), and assuming for the unknown
transversity distribution $h_1^{u/p}$ its upper value, given by saturation of 
the Soffer bound \cite{sof}, $2 |h_{1q}| \le (f_{q/p} + \Delta q)$,
one obtains for the magnitude of 
$A_q^\pi = A_u^{\pi^+} = A_{\bar d}^{\pi^+} = A_d^{\pi^-}$, etc.,
the amazingly large {\it lower} limit \cite{noi3}: 
\bea
&& |A_q^\pi(\langle z \rangle, \langle p_T \rangle)| \>
\gsim \> (0.24 \pm 0.15) \\
&& \langle z \rangle \simeq 0.45 \>, \quad 
\langle p_T \rangle \simeq 0.65 \> \mbox{GeV}/c \nonumber \>.
\label{res}
\vspace*{-12pt}
\eea  
\section{Conclusions}
Single spin asymmetries offer a unique access to new information on
proton structure -- like transversity distributions -- and quark
hadronization -- like the quark analysing power; as such they deserve 
much further experimental and theoretical attention. New data will
soon be available and will help in their understanding and interpretation.

Ideally, one should carefully isolate current quark jets, in processes 
involving transversely polarized protons, and possibly study the $\bfk_\perp$ 
distribution of pions inside them; this would essentially be a direct 
observation of Collins effect and might be feasible at RHIC. The separation
of $z$ dependent Collins function from $x$ dependent transversities
should be possible at future or ongoing DIS experiments \cite{kno}. 

{}From the theoretical point of view, a better understanding of the fundamental
properties of the new spin and $\bfk_\perp$ dependent functions is desirable:
this includes their universality, QCD evolution, factorizability, 
classification and relations among them.     

\section*{Acknowledgments}

M.A. would like to thank the organizers, ECT* and BNL, for
the invitation to a useful and interesting Workshop.

\enlargethispage*{10pt}
 
\nopagebreak

\end{document}